\documentclass[letterpaper]{article} % DO NOT CHANGE THIS
\usepackage{aaai2026}  % DO NOT CHANGE THIS
\usepackage{times}  % DO NOT CHANGE THIS
\usepackage{helvet}  % DO NOT CHANGE THIS
\usepackage{courier}  % DO NOT CHANGE THIS
\usepackage[hyphens]{url}  % DO NOT CHANGE THIS
\usepackage{graphicx} % DO NOT CHANGE THIS
\usepackage{natbib}  % DO NOT CHANGE THIS AND DO NOT ADD ANY OPTIONS TO IT
\usepackage{caption} % DO NOT CHANGE THIS AND DO NOT ADD ANY OPTIONS TO IT
\usepackage{algorithm}
\usepackage{algorithmic}
\usepackage{bm}
\definecolor{bronze}{rgb}{1,1,0.6}
\definecolor{silve}{rgb}{0.969,0.796,0.600}
\definecolor{gold}{rgb}{0.941,0.592,0.600} 
\newcommand{\gold}[1]{\colorbox{gold}{{#1}}}
\newcommand{\silve}[1]{\colorbox{silve}{{#1}}}

\usepackage{amsthm}

\newcommand{\TODO}[1]{\textbf{\color{red}[]}}
\usepackage{newfloat}
\usepackage{listings}
\usepackage{amsmath}
\usepackage{amsfonts}
\usepackage{amssymb}
\DeclareCaptionStyle{ruled}{labelfont=normalfont,labelsep=colon,strut=off} % DO NOT CHANGE THIS
\floatstyle{ruled}
\newfloat{listing}{tb}{lst}{}
\floatname{listing}{Listing}
\title{Deep Inverse Shading: Consistent Albedo and Surface Detail Recovery via Generative Refinement}

\makeatletter
\def\@fnsymbol#1{$\dagger$}
\makeatother

\author {
    Jiacheng Wu\textsuperscript{\rm 1},
    Ruiqi Zhang\textsuperscript{\rm 1},
    Jie Chen\textsuperscript{\rm 1}\thanks{Corresponding author.}
}
\affiliations {
    \textsuperscript{\rm 1}Department of Computer Science, Hong Kong Baptist University, Hong Kong SAR, China\\
    csjcwu@comp.hkbu.edu.hk, csrqzhang@comp.hkbu.edu.hk, chenjie@comp.hkbu.edu.hk\\
}

\begin{document}

\maketitle

\begin{abstract}
Reconstructing human avatars using generative priors is essential for achieving versatile and realistic avatar models. 
Traditional approaches often rely on volumetric representations guided by generative models, but these methods require extensive volumetric rendering queries, leading to slow training. Alternatively, surface-based representations offer faster optimization through differentiable rasterization, yet they are typically limited by vertex count, restricting mesh resolution and scalability when combined with generative priors. Moreover, integrating generative priors into physically based human avatar modeling remains largely unexplored.
To address these challenges, we introduce DIS (Deep Inverse Shading), a unified framework for high-fidelity, relightable avatar reconstruction that incorporates generative priors into a coherent surface representation. DIS centers on a mesh-based model that serves as the target for optimizing both surface and material details. The framework fuses multi-view 2D generative surface normal predictions, rich in detail but often inconsistent, into the central mesh using a normal conversion module. This module converts generative normal outputs into per-triangle surface offsets via differentiable rasterization, enabling the capture of fine geometric details beyond sparse vertex limitations.
Additionally, DIS integrates a de-shading module, informed by generative priors, to recover accurate material properties such as albedo. This module refines albedo predictions by removing baked-in shading and back-propagates reconstruction errors to further optimize the mesh geometry. Through this joint optimization of geometry and material appearance, DIS achieves physically consistent, high-quality reconstructions suitable for accurate relighting.
Our experiments show that DIS delivers SOTA relighting quality, enhanced rendering efficiency, lower memory consumption, and detailed surface reconstruction. 
\end{abstract}


\section{Introduction}
\label{sec:intro}
Reconstructing human avatars is essential for immersive applications such as 3D movies and games. 
Traditional methods focus on recovering geometry from captured data, while recent advances in generative models enable the integration of learned priors, enhancing detail and consistency. This hybrid approach offers a promising direction for building universal, relightable avatars.
% While traditional reconstruction methods focus on recovering geometry from captured data, the recent rise of generative models opens up new opportunities to enhance detail and consistency by integrating learned priors into the reconstruction process, forming a promising hybrid direction for building universal, relightable avatars.

A straightforward approach that builds on generative models involves synthesizing multi-view images and subsequently fusing them into 3D geometry~\cite{long2023wonder3d, voleti2024sv3d, li2025pshumanphotorealisticsingleimage3d}, extending traditional reconstruction pipelines by incorporating learned priors to enrich view synthesis. However, a key limitation of this pipeline is the lack of 3D consistency in the generated views, which leads to challenges in accurately recovering high-fidelity geometries, particularly for detailed human avatars.

Building on this observation, recent methods employ a central 3D volumetric representation guided by 2D generative models. This strategy leverages the rich priors of generative models while maintaining spatial coherence for more accurate and consistent reconstructions. Score Distillation Sampling (SDS) loss~\cite{poole2022dreamfusion, huang2024humannorm, huang2024tech} is widely used in this paradigm, refining volumetric fields with supervision from 2D diffusion models. However, volumetric rendering introduces significant computational overhead and slow training, prompting a growing interest in more efficient alternatives.

% Building on this observation, recent methods employ a central 3D volumetric representation guided by 2D generative models. This approach leverages the detailed priors of generative models while anchoring them to a coherent spatial structure for more accurate and consistent reconstructions. Score Distillation Sampling (SDS) loss \cite{poole2022dreamfusion, huang2024humannorm, huang2024tech} is a widely adopted technique in this paradigm, refining volumetric fields using supervision from 2D diffusion models. However, reliance on volumetric rendering introduces significant computational overhead and slow training, prompting a growing interest in more efficient alternatives.

% To represent geometry for avatar reconstruction, volumetric approaches have been widely explored. 
When it comes to geometry representation for avatar reconstruction, volumetric approaches remain prevalent.
Neural implicit fields~\cite{mildenhall2020nerf}, for example, encode geometry using coordinate-based neural networks, enabling continuous surface representation but requiring dense sampling and incurring high computational costs. Recently, 3D Gaussian Splatting (3DGS)~\cite{kerbl20233d} has improved rendering speed by approximating surfaces using sparse Gaussian primitives. However, it sacrifices geometric precision, which is critical for accurate visibility computation and physically-based rendering (PBR).

To address the computational inefficiency and geometric imprecision of volumetric methods, surface-based representations such as meshes have emerged as promising alternatives for PBR tasks. These representations offer explicit surface definitions, enabling efficient rendering and accurate geometric modeling. Parametric human models (e.g., SMPL~\cite{loper2015smpl} and SMPL-X~\cite{SMPL-X:2019}) offer a deformable mesh structure suitable for animation, but their coarse topology limits capture of high-frequency details essential for photorealistic appearance. Consequently, recent efforts focus on mesh-based strategies that balance computational efficiency with reconstruction fidelity, though challenges remain in optimizing detailed surface variations over sparse mesh topologies.

To address the challenge of modeling fine-grained surface detail on sparse mesh topologies, we propose a surface offset network, inspired by micro-mesh representations~\cite{maggiordomo2023micro, dou2024differentiable}, to capture subtle geometric variations across a dynamic mesh. Our core design is a normal conversion module that maps UV-space surface offsets into 3D normal vectors. This enables gradients from multi-view 2D normal predictions, generated by generative models, to be propagated directly onto the 3D mesh via differentiable rasterization, thereby unifying multi-view signals onto a coherent and optimizable 3D surface. Notably, we represent surface details as surface variations that do not alter the underlying mesh geometry.

While the integration of 2D generative normal introduces valuable fine-grained detail, these predictions often lack angular consistency and exhibit temporal instability. To address this, our method explicitly optimizes the mesh surface against these noisy yet informative generations, enabling the construction of a coherent and high-resolution surface representation. Compared to volumetric approaches, our surface-centric framework provides a more computationally efficient and geometrically precise central representation for physically-based rendering applications.

Nevertheless, integrating generative priors into surface material modeling for PBR in human avatar modeling remains a significant challenge. Existing methods~\cite{kim2024switchlightcodesignphysicsdrivenarchitecture, he2024neurallightrigunlockingaccurate} for general objects often rely on large-scale datasets, but struggle to disentangle shading from intrinsic reflectance properties. To address this, we introduce a prior-driven de-shading module that estimates albedo by removing baked-in shading, leveraging generative shading priors conditioned on geometry and texture. The refined albedo enables reconstruction errors to be back-propagated, allowing for the joint optimization of geometry and appearance. This results in a co-optimization loop that maintains physical consistency across both shape and material.

The contributions are summarized as follows:

\begin{itemize}
\item We propose a mesh-based central surface representation that efficiently reconstructs geometry from multi-view generations. A normal conversion module transforms 2D normal predictions into UV-space surface offsets, enabling gradient flow from 2D to 3D via differentiable rasterization, achieving finer detail than volumetric representations with significantly lower computational cost.

\item We introduce a prior-driven de-shading module that recovers albedo by removing shading, using priors conditioned on surface texture and geometry. This facilitates material disentanglement and captures fine-grained surface detail for physically plausible rendering.

\item We present a unified co-optimization strategy that jointly refines surface geometry and intrinsic appearance. By aligning multi-view consistency with generative priors and coupling geometry with shading in a physically-based rendering loop, our method achieves SOTA relighting accuracy, efficiency, and surface fidelity.
\end{itemize}

% ----------------------------------------------------------------------------------------
\section{Related Works}
\subsection{Volumetric \& Surface Representation}
Volumetric representations~\cite{mildenhall2020nerf, wang2021neus, muller2022instant} model the whole scene and are widely applied in human avatar modeling frameworks. Classical implicit neural fields utilize coordinate-based MLPs to model dynamic human avatars~\cite{gafni2021dynamic, pumarola2021d} by mapping spatial coordinates into implicit neural fields. These methods often combine implicit neural fields with parametric human models to leverage surface constraints for dynamic human avatar modeling~\cite{wang2022arah, zhang2022ndf, peng2024animatable}. However, the extensive querying required for implicit fields leads to long training times and complicates efficient PBR. While 3DGS is efficient to manipulate and apply in human avatar modeling~\cite{qian20233dgsavatar, zhang2024mesh}, it encounters memory challenges as resolution increases, and its sparse nature makes it hard to build a precise surface necessary for accurate PBR calculation. 
In contrast, surface-based approaches focus on capturing surface variations and are commonly used in dynamic human avatar modeling. These methods typically employ parametric models like SMPL~\cite{loper2015smpl} and SMPL-X~\cite{SMPL-X:2019} to improve memory efficiency~\cite{zhang2023explicifying}. However, the limited number of vertices in these models restricts their ability to represent fine surface details, which is crucial for high-quality PBR.
Recent advances introduce micro-mesh representations~\cite{maggiordomo2023micro, dou2024differentiable}, which add displacement to low-polygon meshes as micro-vertices, improving surface detail without significant memory cost. However, effectively applying micro-meshes to dynamic human avatars remains underexplored.
To bridge this gap, we use an offset network to recover micro-vertex displacement, represented as surface variations for dynamic human avatars. We also introduce a normal conversion module that links 2D image features to the 3D mesh surface, resulting in a deformed parametric model with intricate surface variations to be incorporated into PBR for realistic rendering.

\subsection{3D Human Reconstruction with Generative Priors} 
Diffusion models~\cite{rombach2021highresolution} have demonstrated significant potential in image generation~\cite{hu2023animateanyone, zhu2024champ}.
Recent research has investigated their application in multi-view human images, with the goal of fusing these images for avatar geometry reconstruction. While some methods attempt to tackle the multi-view inconsistency by integrating 3D information into the 2D diffusion process~\cite{long2023wonder3d, voleti2024sv3d, li2025pshumanphotorealisticsingleimage3d},
they still struggle with multi-view inconsistencies. This limitation arises because these methods primarily focus on enhancing 2D image consistency rather than addressing the underlying 3D object coherence.
Thus, a more effective strategy is to utilize a base 3D representation and optimize it with priors from 2D diffusion models. This approach promises more accurate and consistent 3D reconstructions for creating high-quality human avatars.
Many current methods employ a pre-trained 2D diffusion model to guide volumetric-centric human representation via SDS loss~\cite{huang2024tech,  xiu2024puzzleavatarassembling3davatars}. However, the extensive sampling point queries slow down training, and the SDS's lack of 3D structure awareness may lead to instability, exemplified by issues like the multi-face Janus problem~\cite{poole2022dreamfusion}.
To address these challenges, our primary goal is to effectively leverage information from prior-driven models to improve surface-centric representation while stably ensuring robust underlying geometry during human avatar reconstruction. 
Specifically, we introduce a normal conversion module that transforms surface variations into normals, allowing us to efficiently distill normal enhancement priors from a pre-trained diffusion model to refine coarse normals during reconstruction. Moreover, we employ inverse shading priors from a prior-driven de-shading module to rectify geometry and material during material learning.
By effectively integrating prior-driven models with human avatar reconstruction, our method accurately models mesh surfaces, disentangles precise material information, and produces more realistic relighting results.

\subsection{Relightable Human Avatar Reconstruction}
Illuminating objects and human avatars involves acquiring novel lighting configurations and understanding material properties~\cite{shu2017neural, guo2019relightables, yang2023towards}.
Traditional methods~\cite{debevec2000acquiring, habermann2019livecap} estimate these properties using photometric stereo techniques under controlled lighting conditions. However, complex real-world illumination poses challenges to these approaches.
To address this limitation, recent advancements have shifted towards learning-based techniques that leverage neural rendering~\cite{li2024uravataruniversalrelightablegaussian, chen2024urhanduniversalrelightablehands} and neural inverse rendering~\cite{srinivasan2021nerv, zhang2021physg, Jin2023TensoIR} to create relightable models grounded in geometric assumptions.
These methods inspire research that decomposes human figures into canonical reflectance fields and incorporates human templates for motion modeling~\cite{chen2022relighting4d, sun2023neural, iqbal2023rana, wu2025fastphysicallybasedneuralexplicit}. Despite their potential, the lack of geometric supervision often leads to inaccuracies in surface reconstruction. More discussions are provided in the extended version.
While some studies~\cite{xu2024relightable, lin2024relightable} enhance the lighting modeling processing under sparse-view inputs, they still rely on the neural implicit representation, which is computationally expensive and unstable. This highlights the need for more efficient methods capable of producing accurate and realistic relightable human avatars.
Other methods address relightable avatar reconstruction in different task settings. For example, MeshAvatar~\cite{chen2024meshavatarlearninghighqualitytriangular} requires dense multi-view inputs and employs a large network for pose-dependent material modeling, leading to long training times and slow inference.
IntrinsicAvatar~\cite{wang2024intrinsicavatarphysicallybasedinverse} uses monocular video and secondary ray tracing on NeRF to achieve detailed surface reconstruction and physically based rendering, but it shares the computational inefficiency typical of implicit methods. Because these approaches focus on different inputs and lack code for sparse-view training, we limit our comparisons to methods designed specifically for sparse inputs~\cite{xu2024relightable, lin2024relightable}.
To improve efficiency, our framework facilitates material disentanglement using an explicit mesh with intricate surface variations. By rectifying geometry and material information guided by a prior-driven model, we achieve precise surface information querying and accurate material disentanglement.

% ----------------------------------------------------------------------------------------
\section{The Proposed Method}
\label{sec: method}
\begin{figure*}[h]
\centering
  \includegraphics[width=1.0\linewidth]{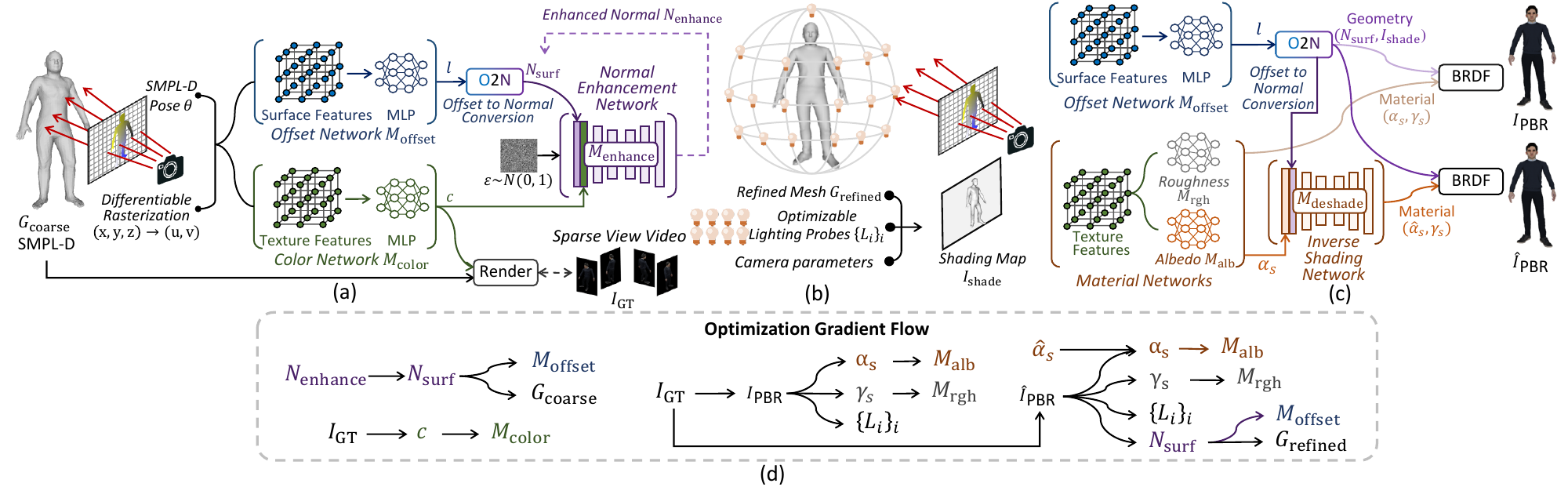 }
  % \vspace{-0.8cm}
    %   \setlength{\abovecaptionskip}{-0.6em}
    % \setlength{\belowcaptionskip}{-1.5em}
\caption{
System diagram:
(a) Prior-driven mesh optimization. The pipeline starts by estimating vertex offsets with the surface offset network $\mathcal{M}_\text{offset}$ to deform a coarse mesh $G_\text{coarse}$. Differentiable rasterization provides pixel UV coordinates, which are converted to surface normals $N_\text{surf}$ via the normal conversion module $O2N$. The color network $\mathcal{M}_\text{color}$ predicts pixel color $c$ and image $I_\text{RGB}$, while the normal enhancement model $\mathcal{M}_\text{enhance}$ refines $N_\text{surf}$ to enhanced normals $N_\text{enhance}$, producing a detailed dynamic mesh $G_\text{refined}$.
(b) Surface-based physically-based rendering. Learnable light probes $\{L_i\}_i$ are placed around $G_\text{refined}$ to estimate visibility. With rasterized surface attributes, we compute albedo $\alpha_s$ via $\mathcal{M}_{\text{alb}}$, roughness $\gamma_s$ via $\mathcal{M}_\text{rgh}$ (both conditioned on shared features from $\mathcal{M}_\text{color}$), and $N_\text{surf}$. These are used in BRDF-based rendering to synthesize the image $I_\text{PBR}$.
(c) Inverse shading and refinement. To remove baked-in lighting from $\alpha_s$, a de-shading module estimates a clean albedo $\hat{\alpha}_s$ conditioned on $\alpha_s$ and $N_\text{surf}$, and computes an improved relit image $\hat{I}_\text{PBR}$.
(d) Joint optimization. Gradients are propagated through the entire pipeline: (1) $\mathcal{M}_\text{color}$ is supervised by $I_\text{RGB}$; (2) $\mathcal{M}_\text{offset}$ and $G_\text{coarse}$ by $N_\text{enhance}$; (3) $\mathcal{M}_{\text{alb}}$, $\mathcal{M}_\text{rgh}$, and $\{L_i\}_i$ by $I_\text{PBR}$; (4) the same modules plus $G_\text{refined}$ by $\hat{I}_\text{PBR}$; and (5) $\mathcal{M}_{\text{alb}}$ is further refined using $\hat{\alpha}_s$.
}
  % \vspace{-0.4cm}
  \label{fig: pipeline}
\end{figure*}

We propose Deep Inverse Shading (DIS), a powerful framework for reconstructing high-fidelity human avatars from sparse-view videos. 
DIS jointly learns detailed geometry and material properties using generative priors within a unified, surface-based representation. 
Built on a dynamic SMPL mesh~\cite{loper2015smpl}, DIS refines geometry via a normal conversion module that maps 2D normal predictions to 3D surfaces, and disentangles shading from albedo with a generative de-shading module for physically plausible appearance modeling, using multi-view generative predictions.

The rest of this section is organized as follows. 
First, we present our prior-driven mesh optimization approach via a normal conversion module, which integrates 2D generative normal priors into 3D meshes using differentiable rasterization. Next, we introduce our physically-based rendering strategy via differentiable surface integration, detailing how material properties are combined with geometry for realistic shading. Finally, we explain our joint material-geometry refinement method through generative inverse shading, which leverages learned priors to co-optimize geometry and appearance in a physically consistent manner.

%-----------------------------------------------------------------------
\subsection{Prior-Driven Mesh Optimization via Generative Normal Conversion}
\label{sec: geometryInitialization}

\begin{figure}[h]
\centering
\includegraphics[width=0.9\linewidth]{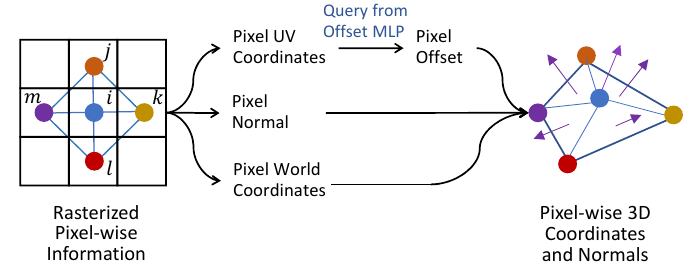}
  \caption{Normal Conversion Module. This module first queries rasterized pixel-level information, including pixel UV coordinates, normals, and world coordinates. 
  Pixel offsets, predicted by an offset MLP, are integrated with the pixel data to compute 3D surface coordinates via Eq.~(\ref{eq: surfWorldCoord}), capturing fine-grained surface variations within each triangle. Finally, point normals are calculated by constructing four surrounding triangles, as described in Eq.~(\ref{eq: normalConversion}).
  }
  \label{fig: normalConversion}
    % \vspace{-0.4cm}
\end{figure}

Differentiable rasterization~\cite{ravi2020pytorch3d} enables gradients to flow from 2D image space to the 3D mesh, allowing effective supervision of 3D geometry from 2D observations. However, traditional parametric models have sparse vertices, limiting their ability to capture fine details. To overcome this, we initialize a coarse mesh $G_\text{coarse}$ and introduce a normal conversion module $O2N$, which refines triangle-level surface normals using an offset network. These offsets are defined in UV space and guided by a prior-driven model conditioned on RGB images and coarse normal predictions.

We first fit an SMPL model to each frame of the input sparse-view videos to obtain a base mesh for deformation. An offset network $\mathcal{M}_\text{offset}$ then predicts scalar offsets $l_\text{v}$ for each SMPL vertex $\mathbf{\hat{x}}_\text{v}$, applied along vertex normals $\mathbf{n}_\text{v}$ to deform the mesh into $G_\text{coarse}$, a coarse approximation of the human surface: 
$\mathbf{\hat{x}}_\text{v} = \mathbf{x}_\text{v} + \mathbf{n}_\text{v} \cdot l_\text{v}.$

Using a differentiable rasterizer, we establish pixel-wise correspondences between image-space coordinates $(x, y)$ and texel-space coordinates $(u, v)$ on the mesh under a given pose ${\theta}$. This mapping enables efficient access to canonical surface attributes at each on-screen pixel. Specifically, for each mesh surface point, we extract interpolated rasterized values such as normal vector $\mathbf{n}\in\mathbb{R}^3$, 3D location $\mathbf{x}$, and query the scalar offset value $l$ from the same offset network $\mathcal{M}_\text{offset}$. In this way, we align the UV coordinates of on-screen pixels, SMPL vertices, and intra-triangle points within a unified UV space. This setup allows us to deform not only the SMPL vertices but also the sub-vertex points inside each triangle to compute 3D surface positions $\mathbf{x_\text{surf}}$:  
{
\footnotesize
\begin{equation}
\mathbf{x_\text{surf}} = \mathbf{x} + \mathbf{n} \cdot l.
\label{eq: surfWorldCoord}
\end{equation}
}
These positions are computed \textit{per-pixel} on the rasterizer for pixel-wise detailed geometric refinement.

The process of normal conversion is illustrated in Fig.~\ref{fig: normalConversion}, where a representative 2D UV-space patch of five rasterized pixel points—denoted as ~$\ m,\ i,\ j,\ k,\ l$ is used to estimate the surface normal at the center point $i$. Their offset values are converted into a coherent normal vector using Eq.(\ref{eq: normalConversion}), capturing the local surface direction variation:
{\scriptsize
\begin{equation}
\begin{split}
\mathbf{n}_i &= 
\mathbf{n}_{\mathbf{\triangle}_{ijk}} + \mathbf{n}_{\mathbf{\triangle}_{ikl}} + \mathbf{n}_{\mathbf{\triangle}_{ilm}} + \mathbf{n}_{\mathbf{\triangle}_{imj}}\\
&= (\mathbf{x}_j - \mathbf{x}_i) \times (\mathbf{x}_k - \mathbf{x}_i) + (\mathbf{x}_k - \mathbf{x}_i) \times (\mathbf{x}_l - \mathbf{x}_i) \\
&+ (\mathbf{x}_l - \mathbf{x}_i) \times (\mathbf{x}_m - \mathbf{x}_i) 
+ (\mathbf{x}_m - \mathbf{x}_i) \times (\mathbf{x}_j - \mathbf{x}_i), \\
\bar{\mathbf{n}}_i &= \frac{\mathbf{n}_i}{\|\mathbf{n}_i\|_2}.
\end{split}
\label{eq: normalConversion}
\end{equation}
}
To obtain the final surface normal map $N_{\text{surf}}$, we integrate the per-pixel surface normal $\bar{\mathbf{n}}_\text{surf}$. In parallel, we query a dynamic color network $\mathcal{M}_\text{color}$ to compute the pixel color $c$ of the coarse mesh $G_\text{coarse}$ under the current pose $\theta$ at each texel coordinate $(u, v)$, integrated as the RGB image $I_\text{RGB}$. $\mathcal{M}_\text{offset}$ and $\mathcal{M}_\text{color}$ each consist of a hash grid encoding followed by a 5‐layer MLP, as in~\cite{muller2022instant}.

To improve both the coarse geometry $G_\text{coarse}$ and the normal map $N_{\text{surf}}$, we introduce a normal enhancement model $\mathcal{M}_{\text{enhance}}\!:\! (N_\text{surf}, I_\text{RGB}) \!\rightarrow\! N_\text{enhance}$ that predicts detailed normals from the surface normal image $N_\text{surf}$ and color image $I_\text{RGB}$. This model provides prior-driven supervision beyond the limited RGB information from sparse‐view video. Efficiently leveraging the graphics pipeline, we perform parallel texel-wise queries for all on-screen pixels. The resulting enhanced normals $N_\text{enhance}$ are used to refine $N_\text{surf}$, and, via the normal conversion module, improvements are back-propagated to optimize both the surface normal information in $\mathcal{M}_{\text{offset}}$, and vertex accuracy of $G_\text{coarse}$ to obtain $G_\text{refined}$.

To implement $\mathcal{M}_{\text{enhance}}$, we adopt Stable Diffusion v1.5 \cite{rombach2021highresolution} as the backbone, pre-trained on a synthetic dataset to generate enhanced normals $N_\text{enhance}$. A modified ControlNeXt~\cite{peng2024controlnext} fuses coarse normals and RGB images as conditional inputs, enabling detailed pixel-level refinement. Rather than generating normals from scratch, the model enhances low-resolution predictions. To handle domain shifts and preserve learned knowledge, only a subset of weights within Stable Diffusion is fine-tuned. The training data, curated with help from a vision foundation model~\cite{khirodkar2024sapiens}, includes coarse normals, RGB images, and high-quality ground truth normals.

\subsection{Physically-Based Rendering via Differentiable Surface Integration}
\label{sec: PBR}
Accurately modeling a physically relightable human avatar requires consideration of illumination, geometry, and surface material properties. 
Central to capturing their interaction is the Bidirectional Reflectance Distribution Function (BRDF)~\cite{walter2007microfacet}, which describes how light reflects off opaque surfaces. The BRDF defines the reflection of incoming light at various angles based on surface attributes, viewing direction, and material properties. The physically-based rendering process is defined in Eq.~(\ref{eq:brdf}):

{
% \footnotesize
\scriptsize
\begin{equation}
\begin{split}
I_\text{PBR}(x_s,\!\mathbf{\omega_o})\!&=\!\int_\Omega\!L_i(\mathbf{\omega_i})\!\cdot\! R_s(x_s,\!\mathbf{\omega_i},\!\mathbf{\omega_o},\!\mathbf{n_s})\!\cdot\!V_s(x_s,\!\mathbf{\omega_i})\cdot\!(\mathbf{\omega_i}\!\cdot\!\mathbf{n_s})d\omega_i, \\
\end{split}
\label{eq:brdf}
\end{equation}
}

where $x_s$ denotes the surface intersection point with the incoming light ray $L_i(\omega_i)$, and $\mathbf{n}_s$ denotes surface normal $\bar{\mathbf{n}}_\text{surf}$. The incoming light ray $L_i(\omega_i)$ is determined by the environment illumination.
The term $R_s(x_s,\mathbf{\omega_i},\mathbf{\omega_o},\mathbf{n_s})$ represents the BRDF, encapsulating surface material information.  
The visibility term $V_s(x_s,\mathbf{\omega_i})$ models occlusions and incidence angles between the lighting source $\{L_i\}_i$ and surface point $x_s$.
These geometric and material components contribute to the physically-based rendering process.

\textbf{Geometry Components.} 
We explicitly model the environment illumination with an array of $16\!\times\!32\!\times3$ learnable light probes $\{L_i\}_i$, uniformly distributed across a surrounding sphere to simulate the environment lighting. 
With the refined dynamic mesh $G_\text{refined}$, we compute visibility via ray-casting on explicit geometry using Open3D~\cite{Zhou2018}, which is typically more efficient than ray-tracing in implicit volumetric fields (e.g.,~\cite{chen2022relighting4d, xu2024relightable}) that require iterative field queries.
And the interaction between lighting and surface normal is assessed by the incidence angle between $\mathbf{\omega_i}$ and $\mathbf{n_s}$. In our case, $\mathbf{n_s}$ is extracted from $N_\text{surf}$.

\textbf{Material Component.} 
We adopt the Micro-facet model~\cite{walter2007microfacet} to simulate the BRDF, with the Cook-Torrance kernel~\cite{cook1982reflectance} to partition surface reflection into diffuse and specular components: 

{
\footnotesize
\begin{equation}
R_s = k_\text{diffuse}\cdot\mathbf{\alpha}_s + k_\text{specular}\cdot\mathbf{\gamma}_s,
\label{eq:cook}
\end{equation}
}

where $k_\text{diffuse}$ and $k_\text{specular}$ denote the fractions of incident light energy allocated to the diffuse and specular components, both set to 1. The diffusive component $\mathbf{\alpha}_s$ is queried from the Albedo Network $\mathcal{M}_{\text{alb}}$, while the specular component $\mathbf{\gamma}_s$ is derived from the Roughness Network $\mathcal{M}_{\text{rgh}}$. Both $\mathcal{M}_{\text{alb}}$ and $\mathcal{M}_{\text{rgh}}$ are decoded from the hash features from $\mathcal{M}_{\text{color}}$. By efficiently combining these components, we compute the final physically-based rendering result $I_\text{PBR}(x_s, \mathbf{\omega_o})$. We then minimize the difference between $I_\text{PBR}$ and $I_\text{GT}$ to optimize light probes and material networks.

\subsection{Joint Material-Geometry Refinement through Generative Inverse Shading}
\label{sec: Rectify}

Once the geometry and material networks produce reliable surface normals and reflectance estimates, we introduce a de-shading module $\mathcal{M}_{\text{deshade}}\!:\! (\bar{\mathbf{n}}_\text{surf}, \alpha_s) \!\rightarrow\! \hat{\alpha}_s$, designed using a three‐level U‐Net structure, to further improve the predicted albedo by removing residual shading. These shading artifacts typically result from inaccuracies in geometry, which lead to lighting being baked into the albedo. 

This module leverages inverse shading priors learned from data introduced in the next section to disentangle intrinsic reflectance from shading. Rather than treating albedo and geometry separately, $\mathcal{M}_{\text{deshade}}$ jointly analyzes the shading-influenced albedo $\alpha_s$ and surface normal $N_\text{surf}$ to estimate a clean, unshaded albedo $\hat{\alpha}_s$. The refined albedo serves both as a physically plausible output and as a learning target for the albedo network $\mathcal{M}_\text{alb}$.
Next, we compute a refined PBR image $\hat{I}_\text{PBR}$ from the unshaded albedo $\hat{\alpha}_s$, replacing the original $I_\text{PBR}$. This refined image is then compared with the ground truth $I_\text{GT}$ to update other components: surface normals via $\mathcal{M}_\text{offset}$, roughness via $\mathcal{M}_\text{rgh}$, and lighting $\{L_i\}_i$. As surface geometry improves, de-shading becomes more reliable, leading to more accurate albedo and further enhancing geometry—a positive feedback loop that refines both shape and appearance jointly. 

% ----------------------------------------------------------------------------------------
\section{Experiments}
\label{sec:experiment}
\subsection{Datasets and Evaluation Metrics}
\label{sec:dataset}
We train and evaluate on eight datasets spanning synthetic and real humans in both indoor and outdoor settings.
For synthetic humans, \textit{SyntheticHuman}~\cite{peng2024animatable} provides 7 dynamic human models with ground truth meshes. We select four subjects and four views to assess mesh reconstruction accuracy, provided in the supplementary materials. Our method achieves the second-best results, while RelightableAvatar(RA)~\cite{xu2024relightable} performs best, as our method primarily refines surface variations rather than altering the underlying topology.
% TODO: Consider move this part to the main text
\textit{SyntheticHuman++}~\cite{xu2024relightable} offers 6 dynamic models with relighting data. We use four individuals and four views for training and comparison with relighting methods.
For real-world scenarios, \textit{People Snapshot}~\cite{alldieck2018video} contains videos of humans moving in circles. We use two outdoor sequences for qualitative evaluation with Relighting4D.
\textit{MobileStage}~\cite{xu2024relightable} includes multi-view images of a person's motion, which we use for training and comparison with RA.
To further analyze our approach via ablation studies, we randomly choose one sample from \textit{ZJUMoCap}~\cite{peng2021neural}, one from \textit{THUman2.0}~\cite{tao2021function4d}, two from \textit{SyntheticHuman++}, and one from \textit{People Snapshot}.
For model training, we create a \textit{Mesh Down-sampling Dataset} with 526 meshes and 35,800 color-normal image pairs, generated with the human vision foundation model Sapiens~\cite{khirodkar2024sapiens}, to train our normal enhancement diffusion model $\mathcal{M}_\text{enhance}$. 
\textit{De-shading Dataset} is constructed using clothing patterns from~\cite{medeirosClothingColorsAndVisualTextures} and synthetic rendering under varied lighting, providing paired data to train our de-shading module $\mathcal{M}_{\text{deshade}}$.
Further details are presented in the supplementary materials.

For surface detail evaluation, we use \textbf{Normal Degree} for normal image cosine similarity. For image quality, we report \textbf{LPIPS$^*$}~\cite{Zhang_2018_CVPR} (LPIPS $\times 10^3$) and \textbf{PSNR}.

{
\begin{table}[hb!]
    \setlength{\abovecaptionskip}{0.4em}
    \begin{center}
        \caption{
        Rendering Memory and Speed Comparison. 
        \gold{Best} and \silve{second-best} results for each metric are highlighted.} 
        \resizebox{0.85\linewidth}{!}{
        \begin{tabular}{c|c|c|c}
            \textbf{Relightable}     & \textbf{Method}  & \textbf{Memory~(GB) $\downarrow$} &  \textbf{Speed (FPS)} $\uparrow$\\
        % \Xhline{1.2pt}
        \hline
         & NDF                     & \silve{5.5}      & \silve{2.48}  \\
       No & AniSDF             & 11.4        & 0.93  \\
        & 3DGS-Avatar         &  \gold{4.0}                  & \gold{50+}              \\
        
        \hline
         & Relighting4D     &  \silve{7.2}       &  0.33       \\
      Yes  & RA                 & 10.89        & \silve{0.52}   \\
        & Ours (DIS)                                           &\gold{3.8}         & \gold{12.16} \\
        \end{tabular}
        \label{table:efficiency}
    }
    \end{center}
\end{table}
}
{
\begin{table}[h!]\footnotesize
    \begin{center}
        \caption{Comparison of Relighting Metrics.
            \gold{Best} and \silve{second-best} results for each metric are highlighted.} 
        \resizebox{1\linewidth}{!}{
        \begin{tabular}{c|c|c|c|c|c|c|c|c|c}
        \multicolumn{1}{c|}{}   &  \multicolumn{1}{c|}{\textbf{Normal}} & \multicolumn{2}{c|}{\textbf{Diffuse Albedo}} & \multicolumn{2}{c|}{\textbf{Visibility}} &  \multicolumn{2}{c|}{\textbf{Relighting}} & \multicolumn{2}{c}{\textbf{Speed}} \\
                & \textbf{Degree} $\downarrow$ & PSNR $\uparrow$ & LPIPS$^*$ $\downarrow$ & PSNR $\uparrow$ & LPIPS$^*$ $\downarrow$ & PSNR $\uparrow$ & LPIPS$^*$ $\downarrow$& Training $\downarrow$ & Rendering $\uparrow$    \\ 
        \hline
        NeRFactor (1 frame) & - & 22.23 &   226.25 & 11.37 &  387 &   21.04  & 313  & 50+ hours & 0.3- FPS\\
        Relighting4D        & 27.41    & \silve{36.16}  &  36.29  &    24.79  & 26.39   &  32.27  & 30.40                &   \silve{40+  hours}      & 0.3- FPS \\
        RA                  & \silve{17.11} & 34.08 & \silve{35.36} & \silve{33.97}  & \silve{21.07} & \silve{34.59}  & \silve{26.21}          & 50+ hours & \silve{0.5- FPS}   \\
        Ours                &  \gold{15.89}   & \gold{36.39} & \gold{34.71}  & \gold{35.11}  & \gold{19.70}  & \gold{35.32} & \gold{25.10} &  \gold{3 hours} & \gold{12+ FPS }\\
        
        \end{tabular}
        \label{table: relighting}
    }
    \end{center}
\end{table}
%\FloatBarrier
}

% ---------------------------------------------------
\subsection{Comparisons on Rendering Efficiency}
\label{sec:RenderingEfficiency}
Images are rendered at $512 \times 512$ on a server with 2 Xeon Silver 4114 CPUs and 1 V100 GPU. Table~\ref{table:efficiency} reports average memory and speed over 100 runs, showing that DIS requires the least memory among all methods by deforming a few vertices and using a network for surface details.
In contrast, implicit approaches such as NDF~\cite{zhang2022ndf} and AniSDF~\cite{peng2024animatable} use coordinate-based networks to cover the entire 3D space, while explicit methods like 3DGS-Avatar~\cite{qian20233dgsavatar} store a dense array of 3D Gaussians, both leading to high memory costs.
DIS also supports PBR and achieves fast rendering via differentiable rasterization, while methods, such as Relighting4D and RA, are constrained by the inefficiency of coordinate-based networks. Although our rendering speed is slower than that of 3DGS methods, our accurate and detailed mesh surfaces offer substantial advantages for surface material learning, particularly when compared to sparse explicit methods that struggle to deliver similar quality.
% ---------------------------------------------------
\subsection{Evaluation on Relighting}
\label{sec: QuantitativeRelighting}

% % \vspace{-2cm}
\begin{figure}[htbp]
    \centering
    % % \resizebox{0.9\linewidth}{!}{
    % \includegraphics[width=0.88\linewidth]{AnonymousSubmission/LaTeX/figures/render_all_v6.5.pdf}
    \includegraphics[width=1\linewidth]{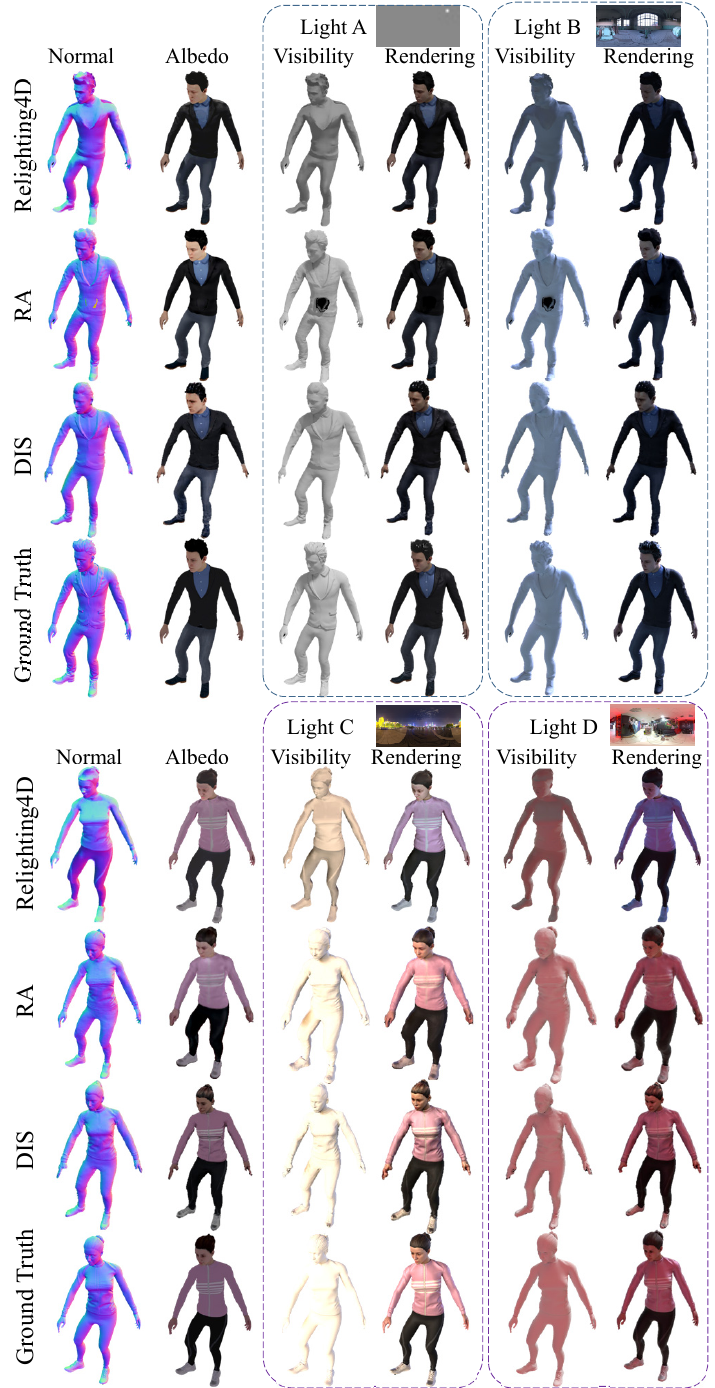}
    
    % }
    \caption{Qualitative comparison on \textit{SyntheticHuman++} across DIS, Relighting4D, and RA. 
    }
    \label{fig:relighting}
    % \vspace{-0.4cm}
\end{figure}

We compare our DIS with leading relighting methods, RA and Relighting4D, and NeRFactor, focusing on static objects trained from a single frame provided in~\cite{xu2024relightable}. Our evaluation results are summarized in Table~\ref{table: relighting} and detailed in the supplementary materials. To resolve scale ambiguity in inverse rendering, we align predicted images to ground truth using channel-wise scale factors~\cite{zhang2021physg}. All metrics are reported for full images and rendering speeds at $512\times512$ resolution on 1 NVIDIA V100 GPU, while RA is evaluated only on foreground regions.
By integrating a refined mesh with detailed surface variations from normal enhancement priors, our method achieves superior normal accuracy and visibility compared to implicit approaches, resulting in the best relighting performance. Our de-shading module further yields the highest albedo metrics.
We also report the shortest training time and fastest rendering speed, exceeding 12 FPS, demonstrating significant efficiency gains. As shown in Fig.~\ref{fig:relighting}, visual comparisons under four novel lighting conditions indicate that Relighting4D often suffers from inaccurate normals and visibility due to unreliable geometry, while RA captures fine details but is prone to instability from implicit fields. In contrast, our method consistently delivers the most stable and realistic results.

\begin{figure}[ht]
    \centering
    \resizebox{1\linewidth}{!}{
    \includegraphics[width=1.0\linewidth]{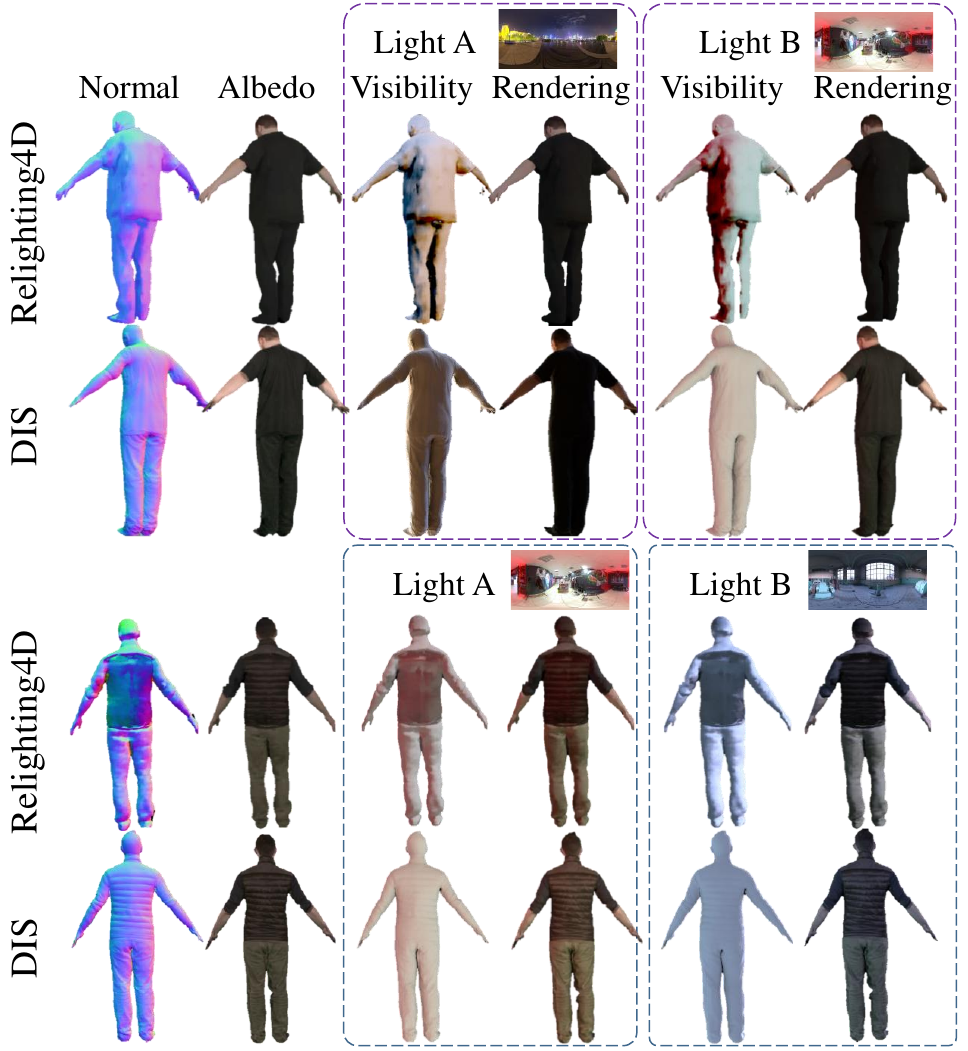}
    }
    \caption{
    Qualitative comparison on \textit{People Snapshot} (real-captured outdoor humans) between DIS and Relighting4D.
    }
    \label{fig:relightingReal}
    % \vspace{-0.4cm}
\end{figure}

To evaluate our method's effectiveness in disentangling real human material attributes, we qualitatively compare with Relighting4D using \textit{People Snapshot}, following their training setup. While RA claims to work well with monocular input, our experiments suggest otherwise, revealing significant challenges in training relightable human avatars in real-world scenarios. In contrast, our method consistently produces good results, aided by normal enhancement priors. As shown in Fig.~\ref{fig:relightingReal}, our mesh-based representation produces superior surface normals and enables more effective material disentanglement than Relighting4D, resulting in more realistic visibility and relighting across diverse lighting conditions. And our albedo is more accurate with less shading effects, benefiting from our inverse shading priors.

\begin{figure}[ht]
    \centering
    \resizebox{1\linewidth}{!}{
    \includegraphics[width=1.0\linewidth]{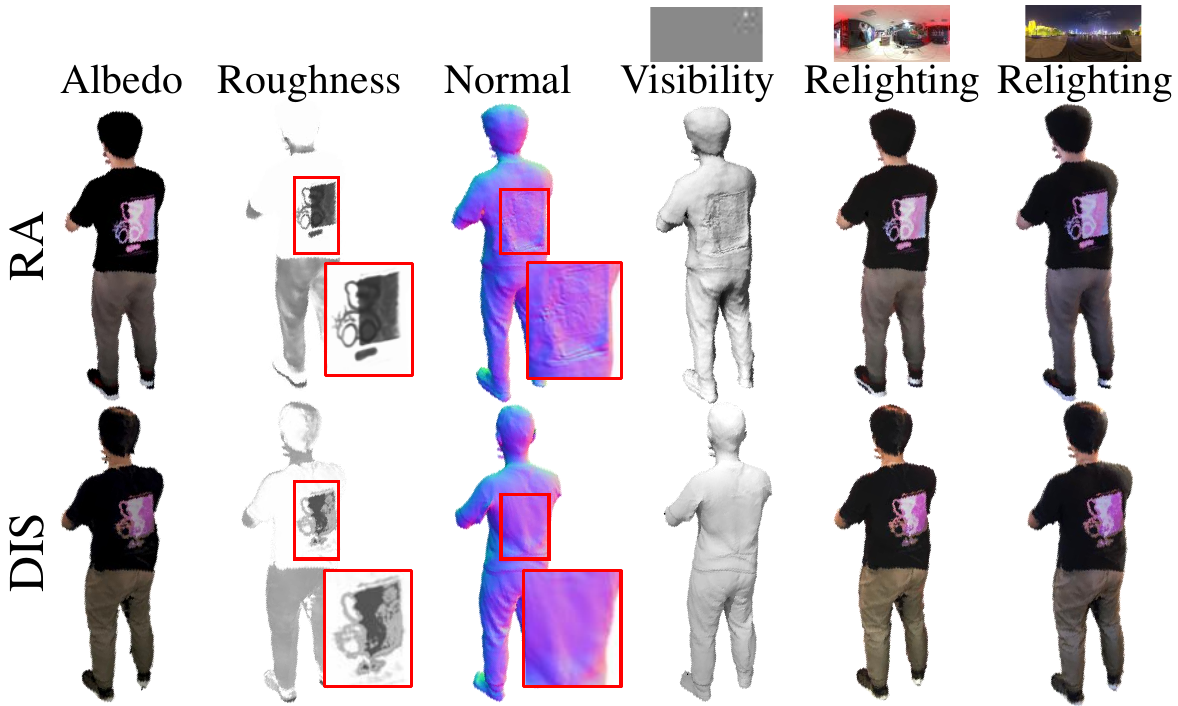}
    }
    \caption{
    Qualitative comparison on \textit{MobileStage} (real-captured indoor human) between DIS and RA.
    }
    \label{fig:relightingRealRA}
    % \vspace{-0.4cm}
\end{figure}

We also conduct qualitative comparisons with RA on \textit{MobileStage}, using their provided model. As shown in Fig.~\ref{fig:relightingRealRA}, our method, benefiting from normal enhancement priors, produces smoother surfaces and more accurate normal maps, resulting in better normal and material disentanglement. For example, the texture on the back of the clothing in our results correctly reflects material differences rather than geometric structure. Our relighting outputs are also more realistic: under the third lighting condition, DIS captures a warm yellow light effect, whereas RA’s output appears blue. 
Since RA does not provide specific training settings, we randomly select 500 frames from eight views for training.

%-------------------------------------------------------------------------
\section{Ablation Studies}
\label{sec:ablation}

We show three ablation studies on two main components: the normal conversion module and the de-shading module.

\subsection{Normal Conversion Module v.s. Mesh Split}
To evaluate the mesh surface expression capability of our normal conversion module $O2N$, we compare our \textit{DIS (w/o $O2N$) \& DIS (w $O2N$)}) with an SMPL‐split‐based approach \textit{Mesh Split} that increases vertex count from 6,890 to 27,554 under identical training. As shown in Fig.~\ref{fig: ablationStudy1}, \textit{DIS (w $O2N$)} captures finer surface details, such as wrinkles in pants, overcoming topology limitations and achieving ultra‐resolution mesh quality. 
Table~\ref{table: ablationStudy1} shows that \textit{Mesh Split} requires three times longer training, while \textit{DIS (w $O2N$)} achieves a significantly lower normal degree, demonstrating effective surface detail for PBR with efficient training.

{
% \vspace{-1em}
\begin{figure}[ht]
    \centering
    \resizebox{0.9\linewidth}{!}{
    \includegraphics[width=1\linewidth]{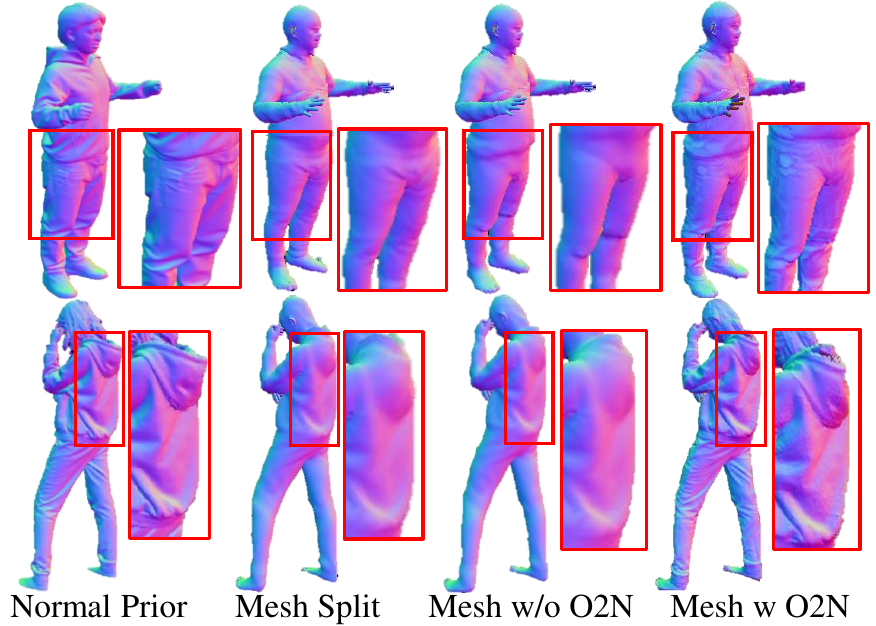}
    }
    \caption{Comparison for Mesh Split with DIS w/wo $O2N$ on \textit{ZJUMocap} and \textit{THUman2.0}.}
    \label{fig: ablationStudy1}
    % \vspace{-0.6cm}
\end{figure} 
}

{
\begin{table}[ht]
\begin{center}
    \caption{Comparisons for Mesh Split and DIS.} 
    \resizebox{0.8\linewidth}{!}{ 
        \begin{tabular}{c|c|c}
     \textbf{Method}  & \textbf{Training Time (Min./Epoch) $\downarrow$} & \textbf{Normal Degree$\downarrow$}\\
    % \Xhline{1.2pt}
    \hline
    \textit{Mesh Split}              & 20+    &  20.36 \\
    \textit{DIS (w/o $O2N$)}   & \textbf{5}   &  24.47 \\
    \textit{DIS (w $O2N$)}   & \textbf{5}   & \textbf{15.34}\\
    \end{tabular}
}
\label{table: ablationStudy1}
% \vspace{-0.5cm}
\end{center}
\end{table}
}

\subsection{PBR w/wo Normal Conversion Module}
\begin{figure}[htbp]
    \centering
    \resizebox{1\linewidth}{!}{
    \includegraphics[width=1\linewidth]{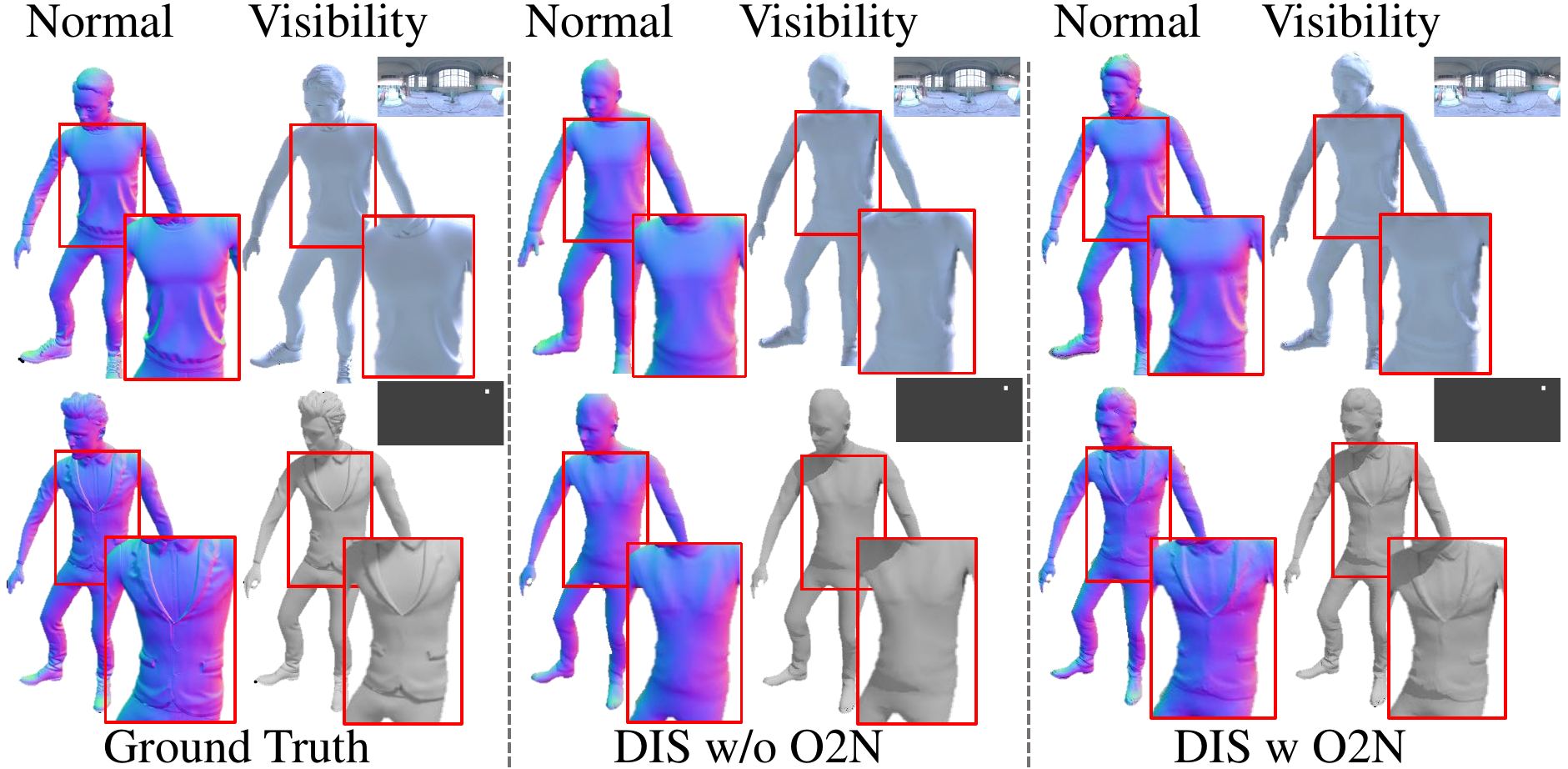}
    }
    \caption{PBR Comparison w/wo $O2N$ on \textit{SyntheticHuman}. }

    \label{fig: ablationStudy_PBR}
    % \vspace{-2em}
\end{figure}

To highlight the importance of the normal conversion module $O2N$ in PBR, we compare visibility results with and without $O2N$ in Fig.~\ref{fig: ablationStudy_PBR}. The red rectangles show that incorporating $O2N$ allows the visibility map to capture fine details, such as wrinkles on the tie and clothing pockets. This also enhances lighting interactions, resulting in sharper edges and rendering results that more closely match GT.

% \vspace{-0.1cm}

\subsection{Albedo \& Normal w/wo De-shading Module}
\begin{figure}[htbp]
    \centering
    \resizebox{1.0\linewidth}{!}{
    \includegraphics[width=1\linewidth]{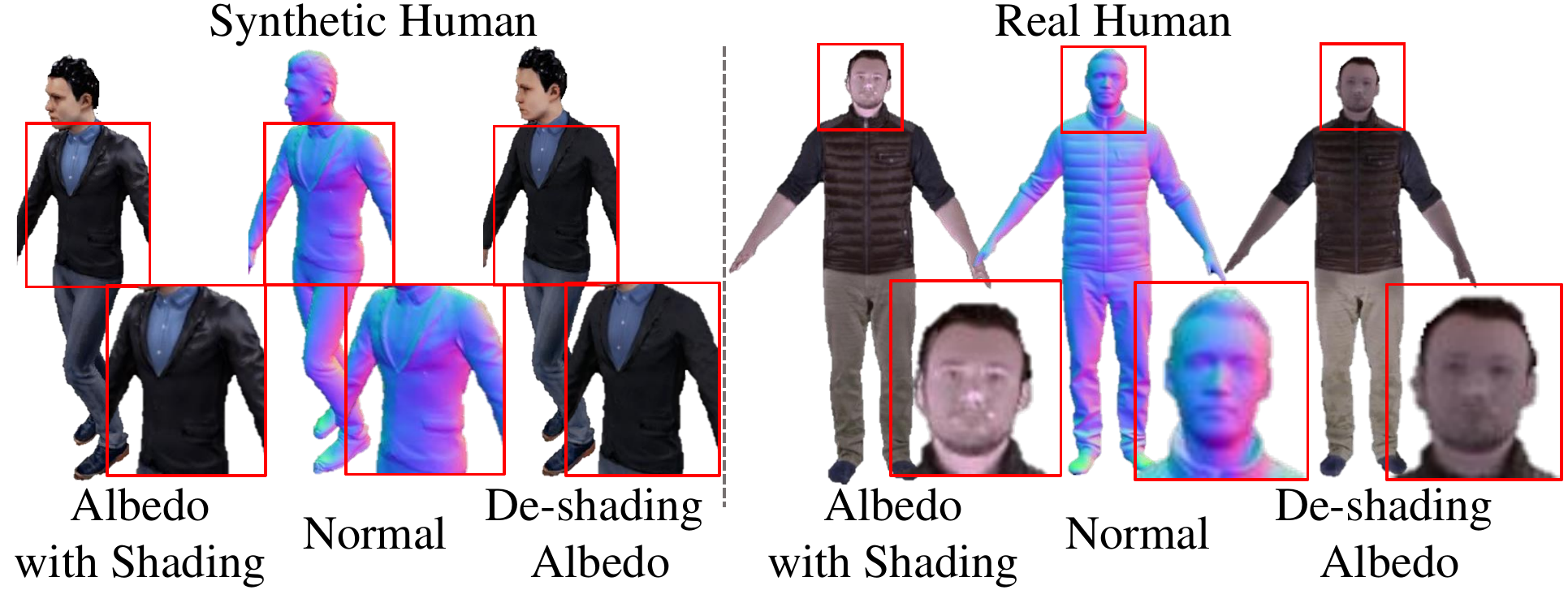}
    }
    \caption{Comparison of Albedo w/wo De-shading Module.}

    \label{fig: ablationDeshading}
    % \vspace{-2em}
\end{figure}

To demonstrate the effectiveness of our de-shading module, Fig.~\ref{fig: ablationDeshading} visualizes the de-shaded albedo for both a synthetic and a real human. As highlighted in red boxes, our de-shading module effectively removes clothing reflections in synthetic humans and facial glare in real humans, demonstrating strong albedo restoration. This capability explains our SOTA albedo accuracy. Additionally, Table~\ref{table: ablationStudyND} shows a decrease in normal degree after applying inverse shading priors, further validating our approach.

% \vspace{-0.1cm}

{
\footnotesize
\begin{table}[htbp]
\begin{center}
    \caption{Surface Normal w/wo De-shading Module.} 
    \setlength{\abovecaptionskip}{-0cm}
    \resizebox{1\linewidth}{!}{ 
    \begin{tabular}{c|c}
    & {\textbf{Normal Degree $\downarrow$}}  \\
    \hline
    Surface Normal w De-shading Module & \textbf{15.89} \\
    Surface Normal w/o De-shading Module & 16.72 \\
    % \Xhline{1.2pt}
    \end{tabular}
}
    \setlength{\belowcaptionskip}{-2cm}

\label{table: ablationStudyND}
% \vspace{-0.5cm}
\end{center}
\end{table}
}

%-------------------------------------------------------------------------
\section{Conclusion}
\label{sec: conclusion}
We introduce DIS for dynamic and relightable avatar modeling, distilling normal enhancement and inverse shading priors onto 3D surfaces through a normal conversion module. Using differential rasterization between 2D images and 3D surfaces, DIS effectively disentangles materials from sparse-view video input, achieving SOTA relighting results and efficient rendering. Joint optimization of generative priors for geometry, material, and surface normals further enhances surface quality and material accuracy.

% ----------------------------------------------------------------------------------------
\section{Acknowledgments}
This research was supported by the Theme-based Research Scheme, Research Grants Council of Hong Kong (T45-205/21-N), and the Guangdong and Hong Kong Universities “1+1+1” Joint Research Collaboration Scheme (2025A0505000003).

\bibliography{aaai2026}

\end{document}

\maketitle

\begin{figure*}[!ht]
\centering
\includegraphics[width=1\linewidth]{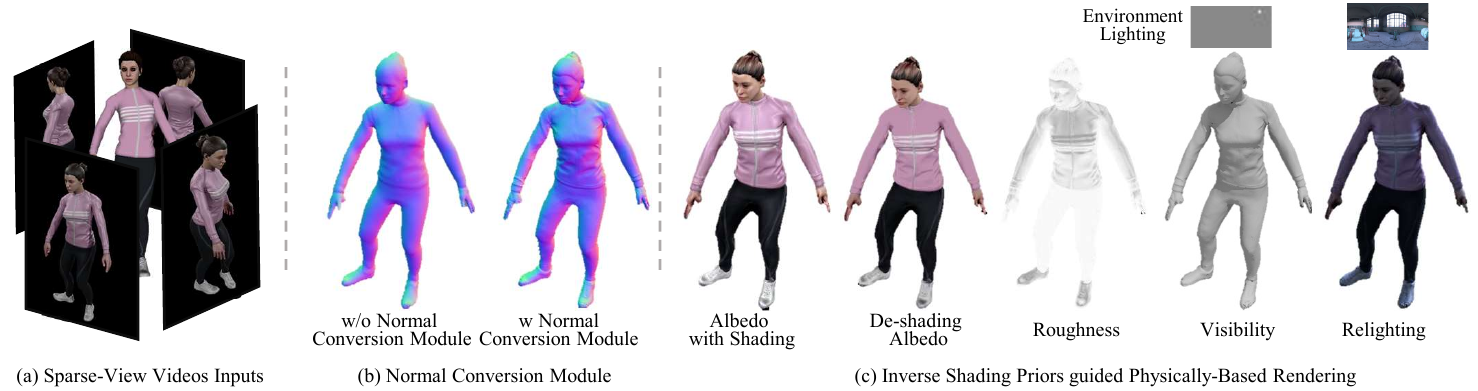}
% \vspace{-0.8cm}
  \caption{Our Deep Inverse Shading (DIS) effectively models physically relightable human avatars using (a) sparse-view video input through several key methods: (b) A normal conversion module is employed to compute surface normals, thereby stabilizing inconsistent normal enhancement priors from a normal enhancement model on a low-polygon mesh. (c) In the physically-based rendering process, the albedo with shading is corrected using inverse shading priors from a de-shading module. This corrected de-shading albedo contributes to generating more accurate PBR results and, inversely, aids in optimizing geometry and material properties. Throughout these processes, we distill consistent albedo and surface details through the integration of generative priors and reconstruction techniques on a 3D surface.
  }
  \label{fig: teaserfigure}
    % \vspace{-0.4cm}
\end{figure*}

In this supplementary document, we will introduce the following contents:
\begin{itemize}
    \item Details of the mesh down-sampling dataset.
    \item Details of the de-shading dataset.
    \item Training details about DIS, normal enhancement diffusion model, and de-shading module.
    \item Detailed Quantitative Comparisons on Mesh Reconstruction and Relighting.
    \item Ablation Studies on the normal conversion module and de-shading module.
    \item Limitation and future direction.

\end{itemize}

\section{Mesh Down-sampling Dataset}
\label{sec: datasetSupp}
This dataset comprises color-normal pairs to train the normal enhancement diffusion model $\mathcal{M}_\text{enhance}$. 
It is constructed using 526 human scans from the THUman2.0 dataset~\cite{tao2021function4d}. 
Each scan is down-sampled into meshes with varying polygon counts using the isotropic remesh algorithm~\cite{hoppe1993mesh} in PyMeshLab~\cite{pymeshlab}, followed by smooth optimization with Pytorch3D, yielding three down‐sampled meshes per scan. 

For training, we generate RGB and normal images at a resolution of $512\times512$ from 20 fixed viewpoints evenly distributed in azimuth from $0^{\circ}$ to $360^{\circ}$ elevation. We also use Sapiens~\cite{khirodkar2024sapiens} to predict normal images from color images in the SyntheticHuman~\cite{peng2024animatable} dataset. The incorporation of a large foundation model ensures robust prior learning in our diffusion model. Ultimately, the dataset consists of 35,800 image pairs.

\section{De-shading Dataset}
\label{sec: deshadingDatasetSupp}
This dataset consists of shading‐albedo‐normal pairs for training the de‐shading module $\mathcal{M}_{\text{deshade}}$. We collect 695 clothing patterns from~\cite{medeirosClothingColorsAndVisualTextures} and 4 synthetic humans from \textit{SyntheticHuman}~\cite{peng2024animatable}. We manually create masks for the upper and lower clothing of each synthetic human, then randomly blend three clothing patterns for each.
For mesh rendering, we generate albedo, albedo with shading, and shading images using a Lambert white texture, applying 26 white lights around the mesh to simulate various lighting conditions. Additionally, we use three camera types with random angles to enrich the dataset. This results in a total of 93,600 pairs for training our de-shading module $\mathcal{M}_{\text{deshade}}$.

\begin{figure*}[h!]
\centering
\includegraphics[width=0.96\linewidth]{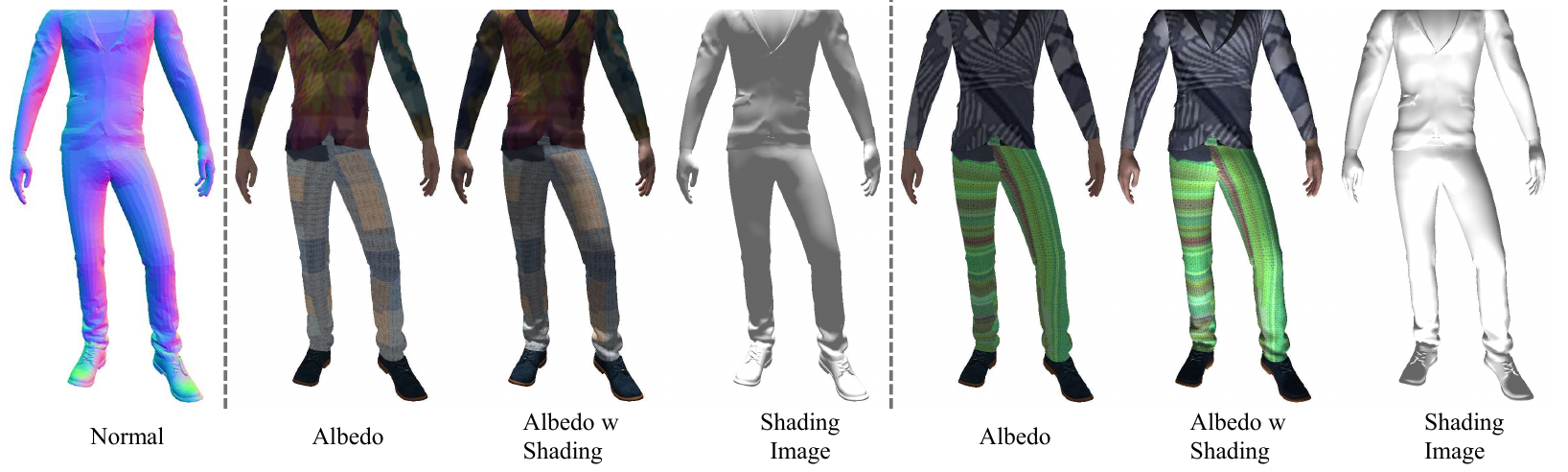}
\caption{Examples of normal maps, albedo maps, albedo maps with shading, and shading images from the De-shading Dataset. The dataset contains four ground-truth meshes, with each mesh rendered under various camera angles and lighting conditions. For each avatar, we render normal maps, albedo maps, albedo maps with shading, and shading images. To increase texture complexity and diversity, clothing patterns from~\cite{medeirosClothingColorsAndVisualTextures} are randomly combined to create different textures based on each avatar's UV islands.}
\label{fig:deshading}
\end{figure*}

\section{Training Details}
\label{sec: trainingSupp}
\subsection{DIS}
\label{sec: NESSupp}

\subsubsection{Network Architecture}
The surface offset network, referred to as $\mathcal{M}_\text{offset}$, takes texel coordinates $(u, v)$ and the pose $\theta$ as inputs to output the offset $l$. 
We employ hash encoding~\cite{muller2022instant} to process the input data, followed by a five‐layer multi‐layer perceptron (MLP). Within this MLP, three fully connected layers employ the SoftPlus activation function, while a residual fully connected layer enhances the ﬁnal output.

The dynamic color network $\mathcal{M}_\text{color}$ is designed to process 2D coordinates $(u, v)$ and pose $\theta$ to output color $c$. 
Similar to $\mathcal{M}_\text{offset}$, this network also consists of ﬁve‐layer MLPs, featuring fully connected layers with ReLU activation functions. The output is subsequently transformed using a Sigmoid activation function.

The albedo and roughness networks $\mathcal{M}_\text{alb}$ and $\mathcal{M}_\text{rgh}$ are structured to process inputs consisting of 2D coordinates $(u, v)$ and pose $\theta$, aiming at producing an albedo item $\mathbf{\alpha}_s$ and a roughness item $\mathbf{\gamma}_s$. 
The hash feature is directly retrieved from $\mathcal{M}_\text{color}$, followed by ﬁve‐layer MLPs composed of five fully connected layers with SoftPlus activation functions. The final outputs are further refined through the application of SoftPlus and Sigmoid activation functions.

\subsubsection{First-stage Training.}
During the initial training phase, we apply differentiable rasterization according to camera parameters to obtain pixel color image $I_\text{RGB}$ and pixel normal image $N_\text{surf}$ converted from the normal conversion module $O2N$. The enhanced normal $N_\text{enhance}$ is generated from $\mathcal{M}_\text{enhance}$ conditioned on $I_\text{RGB}$ and $N_\text{surf}$. At the very beginning of the training stage, we use $I_\text{GT}$ as the color condition. The $\mathcal{M}_\text{offset}$ and $G_\text{coarse}$ are optimized using the mean squared error (MSE) loss and LPIPS loss expressed as follows:

\begin{align}
    L = L_\text{MSE}(N_\text{enhance}, N_\text{surf}) + L_\text{LPIPS}(N_\text{enhance}, N_\text{surf}).
\end{align}

Meanwhile, the color network $\mathcal{M}_\text{color}$ is supervised by the ground truth $I_\text{GT}$ using the MSE loss and LPIPS loss expressed as follows:

\begin{align}
    L = L_\text{MSE}(I_\text{GT}, I_\text{RGB}) + L_\text{LPIPS}(I_\text{GT}, I_\text{RGB}).
\end{align}

We employ the mesh loss $L_\text{mesh}$ to improve the smoothness of the mesh. This composite loss comprises components for mesh edge loss, mesh normal smoothness loss, and mesh Laplace loss, which collectively regularize the mesh structure. For these calculations, we generate the deformed mesh using Pytorch3D~\cite{ravi2020pytorch3d}.
The mesh edge loss penalizes inconsistencies in the lengths of the mesh edges to ensure uniformity. The normal smoothness loss promotes consistency in the normal vectors of adjacent faces, enhancing the natural appearance of the surface. Finally, the mesh Laplace loss facilitates the smoothness of vertex positions by utilizing the Laplacian operator. We define the overall mesh loss as follows:

\begin{align}
    L_\text{mesh} &= \omega_\text{edge} L_\text{edge}\notag
    \\ &+ \omega_\text{normal} L_\text{normal}\notag
    \\ &+ \omega_\text{laplacian} L_\text{laplacian},
\end{align}

\noindent where $\omega_\text{edge}$, $\omega_\text{normal}$, and $\omega_\text{laplacian}$ represent the weights assigned to each component of the loss function.

\subsubsection{Second-stage Training}
In the second training stage, we keep $\mathcal{M}_\text{offset}$ and $\mathcal{M}_\text{color}$ fixed while training the material network $\mathcal{M}_\text{alb}((u, v), \theta) = \mathbf{\alpha}_s$ and $\mathcal{M}_\text{rgh}((u, v), \theta) = \mathbf{\gamma}_s$.
The hash features of $\mathcal{M}_\text{alb}$ and $\mathcal{M}_\text{rgh}$ are obtained from $\mathcal{M}_\text{color}$.
The optimization process focuses on rendering images using pre-defined camera poses and comparing the resulting relighting image $L_o$ ($I_\text{PBR}$) from the rendering equation to the corresponding ground truth $I_\text{GT}$. During the training of these two networks, we utilize both MSE loss and LPIPS loss, defined as follows:

\begin{align}
    L = L_\text{MSE}(I_\text{GT}, I_\text{PBR}) + L_\text{LPIPS}(I_\text{GT}, I_\text{PBR}).
\end{align}

\subsubsection{Third-stage Training}
In the third training stage, all networks except the color network are trained. We utilize the de-shading module to generate inverse shading priors $\hat{\alpha}_s$ conditioned on surface normal $\bar{\mathbf{n}}_\text{surf}$ and albedo with shading $\alpha_s$. This allows for the calculation of a more accurate PBR result $\hat{I}_\text{PBR}$ using $\hat{\alpha}_s$. During this phase, we also adopt MSE loss and LPIPS to optimize all networks. For optimizing the albedo network:

\begin{align}
    L = L_\text{MSE}(\hat{\alpha}_s, \alpha_s) + L_\text{LPIPS}(\hat{\alpha}_s, \alpha_s).
\end{align}

For all other networks:

\begin{align}
    L = L_\text{MSE}(I_\text{GT}, \hat{I}_\text{PBR}) + L_\text{LPIPS}(I_\text{GT}, \hat{I}_\text{PBR}),
\end{align}

\subsection{Normal Enhancement Diffusion Model}
\label{sec: diffusionSupp}
To train the normal enhancement diffusion model, we apply training loss from Stable Diffusion by adding conditions as follows:
\begin{align*}
L_t
&= \mathbb{E}_{t \sim [1, T], \mathbf{x}_0, \boldsymbol{\epsilon}_t} \Big[\|\boldsymbol{\epsilon}_t - \boldsymbol{\epsilon}_\theta(\mathbf{x}_t, t, I_\text{RGB}, N_\text{coarse})\|^2 \Big],
\end{align*}

\noindent where $I_\text{rgb}$ and $N_\text{coarse}$ represent the RGB and the coarse normal images, respectively.

We completely ﬁne‐tune the UNet speciﬁcally for the early conditional injection approach, while only a few small partial weights are adjusted for ControlNeXt. Throughout the entire training period, the VAE remains frozen to maintain its parameters.
This model is trained on 2 NVIDIA V100 GPUs for a total of 60,000 steps, utilizing a learning rate of 1e-5 and a total batch size of 16. The training process is designed to optimize the performance of the UNet in conjunction with the ControlNeXt framework, ensuring that the early conditional injection eﬀectively enhances the model’s capabilities.

\subsection{De-shading Module}
\label{sec: deshadingSupp}
The de‐shading model is structured as a three‐level U‐Net. To train this de‐shading module, we concatenate the normal image and the albedo image with shading as the input to predict the output $I_\text{albedo}$. This approach allows the model to leverage both the geometric information from the normal image and the color information from the albedo image. To ensure that the model focuses on structural similarity during training, we apply the Structural Similarity Index (SSIM) loss to compare the predicted image $I_\text{albedo}$ with the ground truth albedo $\hat{I}_\text{albedo}$. The SSIM loss can be expressed mathematically as follows:

\begin{align*}
L_\text{SSIM} = 1-\text{SSIM}(\hat{I}_\text{albedo}, I_\text{albedo})
\end{align*}

% -------------------------------------------------------------------------------
\section{Detailed Comparisons \& Demonstrations}
\label{sec: detailsSupp}
\subsection{Detailed Quantitative Comparison on Mesh Reconstruction}
For geometry reconstruction evaluation, we use \textbf{Chamfer Distance (CD)} for point cloud similarity, \textbf{Point-to-Surface Distance (P2S)} to calculate distances from points to their nearest reference surface.

\begin{figure}[t]
    \centering
    \includegraphics[width=1.0\linewidth]{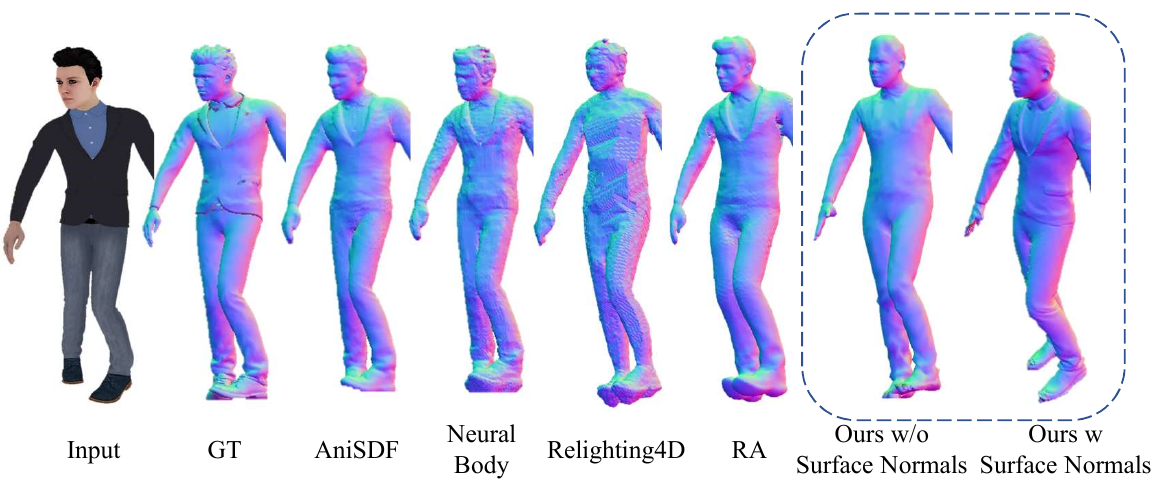}
    \caption{Comparative of surface normals on \textit{Josh}.}
    \label{fig:normal_qual}

\end{figure}

Accurate surfaces with detailed features are crucial for precise material disentanglement. Table~\ref{table:mesh} compares mesh reconstruction quality using P2S and CD metrics on the \textit{SyntheticHuman} dataset. The comparison includes NeuralBody (NB)~\cite{peng2021neural}, AniSDF, Relighting4D, and RA, with average results aggregated for four human models. 
While RA benefits from more precise Signed Distance Fields through its Hierarchical Distance Query Mechanism, it requires over $50^+$ hours of training, compared to just 3 hours for DIS. DIS ranks second in both the P2S and CD metrics, falling short of RA's performance due to inherent topological limitations.
Therefore, we employ a normal conversion module to learn surface normals to address topological limitations. In Fig.~\ref{fig:normal_qual}, we compare qualitative results of the rendered normal maps for the \textit{Josh} model, with our DIS results labeled as 'Ours w/o Surface Normals' and 'Ours w Surface Normals.' The figure illustrates that NB and Relighting4D generate noisy geometries, whereas AniSDF and RA produce detailed surfaces, albeit with some noise due to the marching cube algorithm. In contrast, our learned surface normal incorporates details on the smooth surface, such as fabric pockets. This results in a smoother appearance with fewer artifacts, allowing for a more accurate rendering of complex features compared to the other methods.

{
\begin{table}[ht]
    \begin{center}
    
    \caption{Quantitative Results on Mesh Reconstruction on Synthetic Human Dataset in \textit{P2S/CD}. \gold{Best} and \silve{second-best} results for each metric are highlighted.}
        \resizebox{1.0\linewidth}{!}{
        
        \begin{tabular}{c|c|c|c|c|c}
        P2S$\downarrow$ / CD$\downarrow$ & AniSDF &  NB  & Relighting4D & RA & DIS \\
        \hline
        S1       & 1.29  / 1.21  & 0.62 / 0.65 &  1.67 / 1.58   &\gold{0.53} / \gold{0.54} & \silve{0.54} /\silve{0.54} \\
        % \hline
        
        S2	     & 1.20  / 1.12  & 0.67 / 0.66 &  1.54 / 1.44   &\gold{0.55} / \gold{0.53} & \silve{0.57} /\silve{0.55} \\
        % \hline
        
        S3	     & 1.60  / 1.67  & 0.93 / 1.24 &  1.77 / 1.61   &\gold{0.60} / \gold{0.70} & \silve{0.62} /\silve{0.75} \\
        % \hline
        
        S4       & 0.98  / 1.11  & 0.59 / 0.74 &  1.22 / 1.13   &\gold{0.40} / \gold{0.53} & \silve{0.44} /\silve{0.56} \\
        \hline
        Avg.     & 0.70 / 0.82 & 1.27 / 1.28     & 1.55 / 1.44  &\gold{0.52} / \gold{0.58} & \silve{0.54} / \silve{0.60}\\
        
        \end{tabular}
        \label{table:mesh}
        }
    % \vspace{-0.2cm}
    \end{center}
\end{table}
}

% \vspace{-0.5cm}

\subsection{Detailed Quantitative Comparison on Relighting}
\label{sec: synSupp}
Table~\ref{tab: relighting_all} compares our method with baselines (NeRFactor~\cite{zhang2021nerfactor}, RA~\cite{xu2024relightable} \& Relighting4D~\cite{chen2022relighting4d}) on the \textit{SyntheticHuman++} dataset~\cite{xu2024relightable}, which serves as a comprehensive benchmark for assessing the capabilities of various methods in dynamic relighting scenarios.

{
\begin{table*}[!htbp]\footnotesize
\caption{Detailed Quantitative Results on Four Characters of SyntheticHuman++ Dataset. LPIPS$^*$ denotes LPIPS $\times\ 10^3$. The \gold{best} and \silve{second} results for each experiment data have been highlighted.}
\begin{center}
    \begin{tabular}{c|c|c|c|c|c|c|c|c}
    \multicolumn{1}{c|}{} & \multicolumn{1}{c|}{}   & \multicolumn{1}{c|}{\textbf{Normal}} & \multicolumn{2}{c|}{\textbf{Diffuse Albedo}} & \multicolumn{2}{c|}{\textbf{Visibility}} &  \multicolumn{2}{c}{\textbf{Relighting}}
\\
ID & Method & \textbf{Degree} $\downarrow$ & PSNR $\uparrow$ & LPIPS$^*$  $\downarrow$ & PSNR $\uparrow$ & LPIPS$^*$  $\downarrow$ & PSNR $\uparrow$ & LPIPS$^*$ $\downarrow$\\ 
% \Xhline{1.2pt}
\hline{}
 & NeRFactor (1 frame)   & 33.19  &   17.74        & 201            & 16.12             & 283              & 21.66          & 224 \\           
  jody   & Relighing4D           & 28.47         & \silve{34.02}   & \silve{34.73}   & 24.09             & 29.99            & 30.14          & 31.04               \\
     & RA                    & \silve{15.46}  & 33.03          & {34.91}  & \silve{33.69}     & \gold{22.22}     & \silve{33.41}  & \silve{27.55}       \\
     & Ours                  & \gold{14.67} & \gold{34.42}	& \gold{34.21}	& \gold{34.51}	& \silve{22.26} & \gold{34.76} & \gold{26.07}      \\
\hline{}     
 & NeRFactor (1 frame)   & 35.39 &   29.14        & 167   & 14.85   & 291    & 24.68    & 228 \\ 
  josh   & Relighing4D           & 28.59     &  \silve{35.42}  &  48.98           &  23.67           & 32.16           & 31.53          & 39.71           \\
     & RA                    & \silve{17.31}  & 33.78       & \silve{45.87}   & \silve{32.19}  & \silve{26.36}    & \silve{33.29}          & \silve{32.83}   \\
     & Ours                  & \gold{15.01}   & \gold{35.91}	& \gold{43.19}	& \gold{34.72}	& \gold{21.54} & \gold{34.06} & \gold{30.89}   \\
\hline{}    
 & NeRFactor (1 frame)  & -  &   15.11        & 394   & 7.41  & 538    & 14.06     & 472 \\ 
  megan    & Relighing4D          & 26.73          &  \silve{38.95}  & \silve{14.06}  &   27.94           & 10.72       & 33.86          & 11.79   \\
      & RA                   &  \gold{18.57}  & {36.40}  & {14.09} & \silve{37.01}    & \silve{8.79}       & \silve{35.97}   & \silve{11.61}   \\
      & Ours                 & \silve{17.86}  & \gold{39.02}	& \gold{13.96}	& \gold{37.08}	& \gold{8.56}	& \gold{36.57}	& \gold{11.45}             \\
\hline{}
 & NeRFactor (1 frame) & - &   26.93        & 143   & 7.09   & 435    & 23.77  & 326 \\  
   leonard  & Relighing4D            & 25.85           &  \gold{36.23}  & {47.39}   &  23.46            & 32.70             & 33.54             & 39.06                    \\
     & RA                     &  \silve{17.10}   & 33.09          & \silve{46.57}   & \silve{32.98}      & \silve{26.90}     & \silve{35.69}      & \silve{32.86}    \\
     & Ours                   &  \gold{16.01}   & \silve{36.21}	& \gold{46.49}	& \gold{33.32}	& \gold{26.45} & \gold{35.89} & \gold{31.97}	     \\

    \end{tabular}
    % }
    \end{center}
    \label{tab: relighting_all}
\end{table*}
}

\section{Ablation Study}
To further demonstrate our work's contributions, we present another ablation study focused on the co-optimization mechanism for stabilizing the hallucinated normal $N_\text{enhance}$ as surface normals $N_\text{surf}$.

\subsection{Temporal Consistency}
We evaluate temporal consistency to demonstrate the effectiveness of our co-optimization mechanism in distilling normal mapping details from pre-trained generative models. Specifically, we compare the temporal stability of normal maps produced by four methods: (1) refined mesh $G_\text{refined}$ (\textit{$G_\text{refined}$ Normal}), (2) frame-wise generation using a SOTA normal predictor based on the Stable Diffusion model (\textit{SD Normal})~\cite{ye2024stablenormal}, (3) our context‐guided normal enhancement model (\textit{Hallucinated Normal}), and (4) our final surface normal (\textit{Surface Normal}). We begin by rendering ground‐truth normal maps using models provided in $\textit{SyntheticHuman}$. To establish pixel correspondences between consecutive frames, we employ OpenCV’s \cite{opencv_library} optical flow tracking. Using these correspondences, we perform pixel-wise comparisons across the different normal maps. We then calculate the L1 loss for each pair of consecutive frames and report the average value across all frames in Table~\ref{table:ablation_temporal}. Since pixel correspondences accurately capture the model's motion, any remaining pixel differences are attributed to unstable flickering, making this evaluation an effective measure of temporal consistency. Our results show that \textit{Surface Normal} successfully distills knowledge from the \textit{Hallucinated Normal} while achieving significantly better temporal consistency, outperforming the purely frame-based generative approach (\textit{SD Normal}). Qualitative comparisons, as shown in the supplementary video, further demonstrate that our final surface normal—supported by a robust 3D mesh $G_\text{refined}$—achieves the highest stability without flickering. Additionally, our context-guided normal enhancement model exhibits minimal flickering compared to the other method. Overall, these results highlight the superiority of our approach in maintaining 3D consistency, as evidenced by its outstanding temporal stability.

% \vspace{-0.6em}
\begin{table}[h!]\footnotesize
    \begin{center}
        \caption{Evaluation on Normal Map Temporal Consistency. The \gold{best} and \silve{second} results have been highlighted.}
        \vspace{-0.8em}
        \resizebox{1.0\linewidth}{!}{
        \begin{tabular}{c|c|c|c|c}
        & \textit{$G_\text{refined}$} &  \textit{SD} & \textit{Hallucinated} & \textit{Surface}\\
        ($\times 10^3$) & \textit{Normal} &  \textit{Normal} & \textit{Normal} & \textit{Normal}\\ 
        \hline
        % L1 Loss $\downarrow$ & \gold{49.85} & 79.52 & 78.82 & \silve{51.15} \\ 
        Mean L1 Loss  $\downarrow$ & \gold{1.77} & 2.64 & 2.37 & \silve{1.84} \\ 
        \end{tabular}
        \label{table:ablation_temporal}
    }
    \vspace{-0.3cm}
    \end{center}
\end{table}

% -------------------------------------------------------------------------------
\section{Limitation \& Future Direction}
\label{sec: limitationDirection}
\textbf{Dataset Diversity}
Even though we have created real and synthetic datasets to train the normal enhancement model $\mathcal{M}_\text{enhance}$, our approach still has limitations when it comes to more diverse models. Additionally, even though we have added diversity to the de-shading module $\mathcal{M}_\text{deshade}$ using different textures, our method for removing shadows remains limited for more realistic materials. In the future, we aim to enhance our model's capabilities by creating more real data pairs.

\textbf{Reconstruction Quality of Intricate Regions.}
Due to the fact that the level of detail based on differentiable rasterization optimization depends on the size of the raster, we used a moderate raster size when testing our method. However, for areas that require more detail, such as the face and hands, our optimization capabilities are limited. Therefore, in the future, we will continue to explore the feasibility of using differentiable rasterization to optimize fine regions, further enhancing the capabilities of our method.

% -------------------------------------------------------------------------------

\textbf{Loose Clothes Reconstruction.}
Although the normal conversion module is applied to learn the surface normals, our current approach still relies on a single-layer deformation of the SMPL model, which is inadequate in certain situations, such as when dealing with loose clothing. Future research could explore the possibility of learning surface normals across different layers and developing more advanced parametric clothed human models that accurately represent these scenarios. By incorporating our normal conversion module, these models could enhance the depiction of intricate details in clothing dynamics, even with low-polygon meshes.

\textbf{Head \& Hand Reconstruction.}
Some artifacts persist in reconstructing the head and fingers, primarily due to the inherent topology and fitting limitations of the SMPL model, which affects most SMPL-based approaches. Recent 3DGS frameworks have shown improvements in head modeling, and advanced fitting algorithms offer promising enhancements for finger reconstruction. We plan to explore hybrid representations in future work to address these limitations.

\textbf{Depth Discontinuity.}
Our training pipeline ensures depth continuity, addressing challenges related to depth discontinuities observed in the O2N approach. The normal enhancement network, which is conditioned on RGB inputs, effectively detects depth boundaries and corrects associated errors during optimization. Moreover, non-connected triangles occur only within individual rasterization cycles to assist geometry optimization but are excluded from the final reconstructed model. We plan to refine the O2N module in future work to fully resolve this issue.

% -------------------------------------------------------------------------------
\bibliography{aaai2026}